\documentclass[aps,prx,reprint,amsmath,amssymb,floatfix]{revtex4-2}
\usepackage[bookmarks=false,linkcolor=blue,urlcolor=blue,colorlinks,citecolor=blue]{hyperref}
\usepackage{graphicx}
\usepackage{xcolor}

\newcommand{\figref}[2]{\hyperref[#1]{\ref{#1}(#2)}}

\usepackage{color}

\newcommand{\normord}[1]{:\mathrel{#1}:}

\newcommand{\be}{\begin{equation}}
\newcommand{\ee}{\end{equation}}
\newcommand{\bea}{\begin{eqnarray}}
\newcommand{\eea}{\end{eqnarray}}

\newcommand{\lp}{\left(}
\newcommand{\rp}{\right)}

\newcommand{\Avg}[1]{{\left\langle #1\right\rangle}}

\begin{document}
\title{Correlated insulating states in carbon nanotubes controlled by magnetic field}
\author{Assaf Voliovich$^1$, Mark S. Rudner$^{2,3}$, Yuval Oreg$^1$, and Erez Berg$^1$}
\date{\today}
\affiliation{$^1$Department of Condensed Matter Physics, Weizmann Institute of Science, Rehovot, 76100, Israel\\
$^2$Department of Physics, University of Washington, Seattle, WA 98195-1560, USA\\
$^3$Niels Bohr Institute, University of Copenhagen, 2100 Copenhagen, Denmark}
\begin{abstract}

We investigate competing insulating phases in nearly metallic zigzag carbon nanotubes, under conditions where an applied magnetic flux approximately closes the single particle gap in one valley. 
Recent experiments have shown that an energy gap persists throughout magnetic field sweeps where the single-particle picture predicts that the gap should close and reopen.
Using a bosonic low-energy effective theory to describe the interplay between electron-electron interactions, spin-orbit coupling, and magnetic field, we obtain a phase diagram consisting of several competing insulating phases that can form in the vicinity of the single-particle gap closing point.
We characterize these phases in terms of spin-resolved charge polarization densities, each of which can independently take one of two possible values consistent with the mirror symmetry of the system, or can take an intermediate value through a spontaneous mirror symmetry breaking transition. 
In the mirror symmetry breaking phase, adiabatic changes of the orbital magnetic flux drive charge and spin currents along the nanotube.
 We discuss the relevance of these results to recent and future experiments.
\end{abstract}
\maketitle

\section{Introduction}

Carbon nanotubes (CNTs) offer a versatile platform for studying the exotic quantum many-body physics of one-dimensional (1D) electronic systems.
A variety of intriguing phenomena have been observed, including Wigner crystal formation~\cite{Shapir2019} and possible Luttinger liquid behavior~\cite{Bockrath1999}, strong~\cite{Leturcq2009} and spatially-resolved~\cite{Benyamini2014} electron-phonon coupling~\cite{SuzuuraAndo2002, MarianiVonOppen2009}, and highly efficient multiple carrier generation through photoexcitation cascade~\cite{Gabor2009}.
Despite the extremely weak spin-orbit coupling (SOC) in graphene, the parent material for CNTs, remarkably strong SOC has been observed~\cite{IlaniKuemmeth2008, Steele2013}  and exploited for spin qubit operation~\cite{Flensberg2010, Weiss2010, Laird2013} while also providing means for efficient spin relaxation~\cite{BulaevTrauzettelLoss2008, Borysenko2008, Churchill2009, RudnerRashba2010} and nanomechanical coupling~\cite{Palyi2012}.
\begin{figure}[ht!]
    \includegraphics[width=0.95\columnwidth]{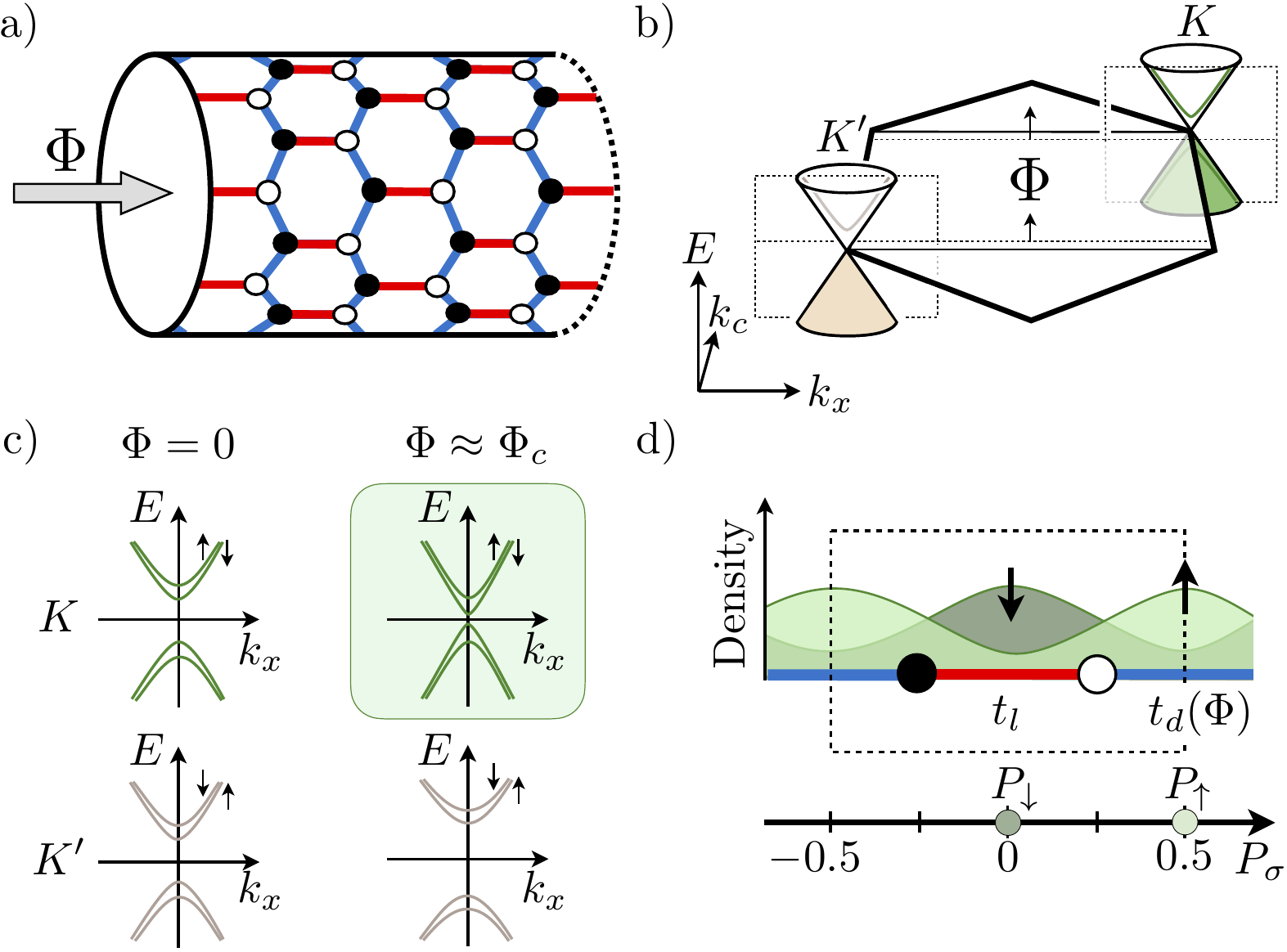}
    \caption{Competing insulating phases of a nearly metallic zigzag carbon nanotube (CNT).
    a) Geometry of the CNT. A magnetic flux $\Phi$ threaded through the CNT tunes the one-dimensional electronic sub-band dispersion (see panels b and c).
    Filled (empty) circles indicate $A$ ($B$) sites.
    b) Two-dimensional graphene Brillouin zone and low-energy dispersion, with the lowest one-dimensional sub-band sections of the CNT indicated.
    While geometric quantization of the circumferential wave number $k_c$ would suggest that the CNT should be metallic, curvature of the CNT offsets the subbands from the Dirac points (dashed lines), leading to a small single-particle gap that can be controlled by the flux, $\Phi$.
    c) Lowest sub-band dispersion relations in valleys $K$ and~$K'$, at zero flux ($\Phi = 0$) and near the critical flux ($\Phi \approx \Phi_c$).
        Our analysis focuses on the effective one-dimensional system comprised of the low energy modes for $\Phi \approx \Phi_c$, highlighted by the green box. (Here we assume that the zero flux gap due to curvature/strain is larger than the SOC strength.)
    d) Distinct insulating phases are characterized by the spin-resolved charge polarization densities, $P_\uparrow$ and $P_{\downarrow}$, shown here 
    with a relative offset of one half unit cell as in the phase realized at large spin-orbit coupling, $\lambda$, see Fig.~\ref{fig:phase_diagram}(a).
     \label{fig:setup}}
     \vspace{-0.24in}
 \end{figure}
 
At the single-particle level, the electronic structure of a CNT is determined by its diameter, $d$, and chirality -- i.e., the orientation of the honeycomb lattice of carbon atoms relative to the nanotube axis~\cite{LairdRMP}.
Here we focus on nominally metallic zigzag CNTs, as illustrated in Fig.~\ref{fig:setup}a.
In such CNTs the combination of quantization of the circumferential wave vector, $k_c$, and curvature-induced strain yield low energy one-dimensional subbands with small gaps that decay systematically with CNT diameter as $\sim 1/d^2$ (see Fig.~\ref{fig:setup}b and Refs.~\cite{KaneMele1997, Lieber2001}).
By applying a magnetic flux, $\Phi$, through the nanotube, these 1D subbands can be shifted relative to the Dirac points of the underlying two-dimensional graphene dispersion, in principle allowing the gap to be closed in one valley at a critical value of the applied flux, $\Phi_c$ (Figs.~\ref{fig:setup}b,\ref{fig:setup}c).

Intriguingly, experiments on ultraclean, suspended small-gap CNTs have shown that while the gap can be tuned by applied flux, the minimal gap achieved during continuous sweeps (where the gap first shrinks and then grows again) generically retains a significant nonzero value (of order a few tens of meV)~\cite{Bockrath2009}.
This failure of the single-particle picture to account for the observed behavior suggests that electron-electron interactions (which are essentially unscreened at charge neutrality in suspended devices) may dramatically transform the nature of the ground state in this regime. Theoretical studies have predicted various interaction-driven insulating states in carbon nanotubes~\cite{Egger1997,Balents1997,Krotov1997,Bunder2008,Rontani2014,varsano2017carbon}.

In this work we investigate the effects of electron-electron interactions on the ground states of nominally metallic carbon nanotubes.
We find that the system obtains one of several competing insulating ground states, controlled by the interplay between spin-orbit coupling, magnetic field, and interactions.
We characterize these states in terms of spin-resolved charge polarization densities, $P_\sigma$. Here, $\sigma =\, \uparrow,\downarrow$ indicates the spin direction relative to the quantization axis along the nanotube, which we denote by $\hat{x}$. The spin-resolved polarization densities are illustrated schematically in Fig.~\ref{fig:setup}d. 
In preserving the mirror symmetry of the zigzag nanotube (reflection across a mirror plane $M_x$ that bisects the longitudinal bonds shown in red in Fig.~\ref{fig:setup}a), each polarization component can obtain one of two possible values, $0$ or~$1/2$. In the absence of interactions, these two cases correspond to Wannier centers for spin-up and spin-down electrons positioned on the longitudinal (red) or ``diagonal'' (blue) bonds of the CNT shown in Fig.~\ref{fig:setup}a.
We further show that under certain conditions, states which spontaneously break the mirror symmetry may be favored, where $P_{\uparrow,\downarrow}$ are neither $0$ nor $1/2$.   

\section{Physical picture}
\label{se:physical-picture}
Before beginning our detailed analysis, we give a simple physical picture that helps to motivate our results.
As discussed for example in the supplementary material of Ref.~\cite{Efroni2017} and illustrated in Figs.~\ref{fig:setup}a) and ~\ref{fig:setup}d), the single-particle electronic states in one valley of a zigzag CNT, for each spin species, can be described via a mapping to the Su-Schrieffer-Heeger (SSH) model of polyacetylene~\cite{SSH}.
In Fig.~\ref{fig:setup}a we identify two types of bonds: ``longitudinal bonds'' that run parallel to the nanotube axis (shown in red color) and ``diagonal bonds'' which have components oriented 
around its circumference (shown in blue).
The low energy electronic states in the $K$ and $K'$ valleys are described by large values of the circumferential wave vector component, $k_c$ (Fig.~\ref{fig:setup}b).

Fixing $k_c$ to the value corresponding to $K$ or $K'$ yields an effective 1D 
tight binding model (Fig.~\ref{fig:setup}d) with alternating hopping amplitudes $t_\ell$ and $t_d$ arising from the longitudinal and diagonal bonds, respectively.
For a flat sheet of graphene and in the absence of a magnetic field, the hopping amplitudes are identical along all bonds. 
For the CNT, the combination of phases arising from circumferential motion ($k_c \neq 0$), curvature-induced strain, magnetic flux, and spin-orbit coupling affects the interference between amplitudes for hopping along the two diagonal bonds entering each site; in particular, this allows the {\it magnitude} of the hopping amplitude $t_d(\Phi)$ in the effective 1D model to be tuned by the flux $\Phi$.

Using the flux-tunability of $t_d(\Phi)$, various regimes of effective dimerization can be explored. 
For a noninteracting system these correspond to cases where the Wannier centers for electrons in the low energy spin-up and spin-down bands can independently be centered either on the longitudinal or the diagonal bonds. 
In a many-body setting at charge neutrality, these situations correspond to insulating phases with charge polarization densities $P_\sigma$ for spin-up ($\sigma =\ \uparrow$) and spin-down ($\sigma = \ \downarrow$) electrons taking values 0 or $1/2$ (in units of 
the electron charge,~$e$), see Fig.~\ref{fig:setup}d.   
With interactions there is an additional possibility of spontaneous mirror symmetry breaking, where at least one of the $P_\sigma$'s takes a value different from $0$ and $1/2$.
Moreover, interactions can shift phase boundaries, and renormalize energy gaps, the spin-orbit coupling, and the orbital moments in each valley. 
Below we treat the interacting problem in detail, and map out the resulting phase diagram of the system (see Fig.~\ref{fig:phase_diagram}).

\section{Model}
To analyze the low-energy behavior of the system, we use a one-dimensional continuum model. The model describes the (spinful) electrons in the lowest energy sub-band of a single valley, in which the single-particle gap is nearly closed by an applied magnetic flux (shaded panel in Fig.~\ref{fig:setup}c).
Within this model the Hamiltonian takes the form (with $\hbar = 1$ throughout, unless otherwise noted):
     \begin{eqnarray}
         {\cal H} &=& \int dx \Big[   -v  \sum_{\sigma, r}   r\psi^\dagger_{r \sigma} i\partial_x  \psi_{r \sigma}\!-\! \sum_{\sigma,r} \left(h \sigma +\mu \right) \rho_{r\sigma} \nonumber\\ 
          &+& \sum_{\sigma} \frac{2v}{d} \left(\lambda \sigma - f \right) \left(i\psi^\dagger_{R \sigma} \psi_{L \sigma} + {\rm h.c.}\right)\Big] + \mathcal{H}_{\rm int},  
          \label{eq:SpinFullHamitonian}
     \end{eqnarray}
     where $v$ is the Fermi velocity, $d$ is the nanotube's diameter, $\psi^\dagger_{r \sigma}$ $(\psi_{r \sigma})$ is the creation (annihilation) operator of a left ($r=L$) or right ($r=R$) moving electron with spin $\sigma =\ \uparrow, \downarrow$ at position $x$,  $\rho_{r \sigma} =\psi^\dagger_{r \sigma} \psi_{ r \sigma}$, and $\mathcal{H}_{\rm int}$ describes the electron-electron interaction, to be defined below. 
     (When it is not used as a subscript, we assign~$\sigma=+1$ for spin up and $\sigma = -1$ for spin down; similarly, we let~$r=+$ for right movers and~$r=-$ for left movers and~$\bar{r}=-r$.)
     The coefficient $\frac{2v}{d}(\lambda \sigma - f)$ is the spin-dependent mass arising due to a combination of spin-orbit coupling, $\lambda$, and the normalized orbital flux, $f = \frac{\Phi - \Phi_c}{2\pi\hbar/e}$.  
     The Zeeman energy due to the application of the axial field is denoted by $h$, and $\mu$ is the chemical potential. 
     
     The CNT is invariant under a mirror reflection $M_x$ with respect to a plane perpendicular to the nanotube axis that passes through the middle of the longitudinal bonds (red color in Fig.~\ref{fig:setup}). 
     To see how this symmetry is manifested in our description, it is helpful to recall that, microscopically, the right- and left-mover single particle states have a pseudospin structure described by particular sets of amplitudes on the $A$ and $B$ sublattices of carbon atoms.
     Specifically, due to the zigzag CNT's geometry and for our choice of basis, these amplitudes correspond to the eigenstates of the second (i.e., $y$) Pauli matrix in the sublattice space; consequently, electronic states supported on the $A$ and $B$ sublattices are created by the operators $\psi_{A\sigma}^\dagger = \frac{1}{\sqrt{2}}(\psi^\dagger_{R\sigma} + \psi^\dagger_{L\sigma})$, $\psi_{B\sigma}^\dagger = -\frac{1}{i \sqrt{2}}(\psi^\dagger_{R\sigma} - \psi^\dagger_{L\sigma})$, respectively. 
     Noting that the $A$ and $B$ sublattices are interchanged by the mirror operation $M_x$, $\psi^\dagger_A \leftrightarrow \psi^\dagger_B$, the relations above determine that this symmetry acts as $\psi_{R\sigma}\rightarrow i\psi_{L\sigma}$, $\psi_{L\sigma}\rightarrow -i\psi_{R\sigma}$. 
     It is straightforward to check that all terms in Eq.~(\ref{eq:SpinFullHamitonian}) are invariant under this symmetry.
    
    We describe the electron-electron Coulomb interaction, which we assume to be screened by a nearby metallic gate, via 
    \begin{equation}
        \mathcal{H}_{\rm{int}} = \frac12 \sum_{\alpha\beta}\int dx\int dx' U_{\alpha\beta}(x-x') \normord{\rho_{\alpha}(x) \rho_{\beta}(x')},
    \end{equation}
    where $\alpha,\beta=\{A,B\}$ are sublattice indices, $\rho_{\alpha}(x) = \sum_\sigma \psi^\dagger_{\alpha,\sigma}\psi_{\alpha,\sigma}$, and $\normord{O}$ denotes normal ordering of the operator $O$ relative to the (non-interacting) Fermi sea.
    Here $U_{AA}(x-x')=U_{BB}(x-x')$ describes the interaction between two electrons on the same sublattice, whereas $U_{AB}(x-x')=U_{BA}(x-x')$ is the interaction between electrons on different sublattices. 
     
     To treat the interacting system, we bosonize Hamiltonian in Eq.~(\ref{eq:SpinFullHamitonian}) following the standard procedure~\cite{Giamarchi2003}.
     We first introduce canonical bosonic fields $\theta_\sigma$, $\phi_\sigma$, satisfying the commutation relations  $[\phi_{\sigma}(x), \theta_{\sigma'}(x')]=i\pi\delta_{\sigma\sigma'}\Theta(x'-x) + i\pi (1-\delta_{\sigma\sigma'})$. 
     The electronic field operator is written as $\psi_{r\sigma}=\frac{1}{\sqrt{2\pi d}}e^{i(\theta_{\sigma}+r\phi_{\sigma})}$, and the density of the electrons with spin $\sigma$ is given by 
     \begin{equation}
     \rho_{\sigma}=\frac{1}{\pi}\partial_{x}\phi_{\sigma}.
     \label{eq:rhos}
     \end{equation}
We decompose the Hamiltonian into four contributions:
\begin{equation}
H=H_{0}+H_{{\rm Z}}+H_{\lambda,f}+H_{{\rm BS}},
\label{eq:Hbosonized}    
\end{equation}
where $H_0$ captures the electronic kinetic energy (neglecting spin-orbit coupling and orbital coupling to the applied magnetic flux) as well as the forward scattering part of the electron-electron interaction, $H_{\rm Z}$ describes the Zeeman coupling and chemical potential, $H_{\lambda,f}$ captures the spin-orbit and orbital magnetic couplings, and $H_{\rm BS}$ captures the backscattering part of the electron-electron interaction. 

For analyzing the bosonized Hamiltonian, it is convenient to introduce spin ($s$) and charge ($c$) fields: $\phi_{j}=\frac{1}{\sqrt{2}}\left(\phi_{\uparrow}+\eta_{j}\phi_{\downarrow}\right)$, $\theta_{j}=\frac{1}{\sqrt{2}}\left(\theta_{\uparrow}+\eta_{j}\theta_{\downarrow}\right)$, where $j=c,s$ and $\eta_{c}=+1$, $\eta_{s}=-1$. 
In terms of these new fields, we have~\cite{Giamarchi2003}
\begin{equation}
H_{0}=\sum_{j=c,s}\frac{v_{j}}{2\pi}\int dx\left[K_{j}(\partial_{x}\theta_{j})^{2}+\frac{1}{K_{j}}(\partial_{x}\phi_{j})^{2}\right],
\label{eq:H0bosonized}
\end{equation}
with $v_{j}=v\left(1+\frac{U_{+}(1+\eta_{j})/2\, +\, U_{-}\eta_{j}}{\pi v}\right)^{1/2}$ and $K_{j}=v/v_{j}$, where $U_\pm = \frac12 \int dx(U_{AA} \pm U_{AB})$. The interaction $U_+$ corresponds to forward-scattering processes, and is typically larger than $U_-$ by a factor of the order of $d/a$, where $a$ is the lattice spacing~\cite{Kane1997Coulomb}. 
The Zeeman and chemical potential terms are given by
\begin{equation}
H_{\rm{Z}} = -\int dx\left(\mu\frac{\sqrt{2}}{\pi}\partial_{x}\phi_{c}+h\frac{\sqrt{2}}{\pi}\partial_{x}\phi_{s}\right),
\label{eq:Hz}
\end{equation}
while the mass terms due to spin-orbit coupling and the magnetic flux are written as:
\begin{eqnarray}
    H_{\lambda,f} = -\frac{2v}{\pi d^{2}}\int dx\Big[&\lambda&\sin\left(\sqrt{2}\phi_{s}\right)\sin\left(\sqrt{2}\phi_{c}\right) \nonumber \\ +&f&\cos\left(\sqrt{2}\phi_{s}\right)\cos\left(\sqrt{2}\phi_{c}\right)\Big].
    \label{eq:lambdaf}
\end{eqnarray}
Finally, the backscattering interaction terms are given by
\begin{equation}
    H_{{\rm BS}}=\frac{v}{\pi d^{2}}\int dx\left[g_{s}\cos(2\sqrt{2}\phi_{s})-g_{c}\cos(2\sqrt{2}\phi_{c})\right],
    \label{eq:BS}
\end{equation}
where $g_{s}=g_{c}=\frac{U_-}{2\pi v}>0$. 
In the following, we study the insulating ground states realized in this model as~$\lambda$, $f$, $h$, and $\mu$ are tuned.

\section{Polarization}

The insulating states of the nanotube are most conveniently characterized by the values of the electric polarizations~\cite{King-SmithVanderbilt1993, RestaRMP1994, RestaPRL1998, MarzariRMP2012, Asboth2016} of spin up and spin down electrons. 
A boundary between two states with different polarizations hosts a fractional charge and/or fractional spin, equal to the difference of the polarization densities in the two states. 

Our goal in this section is to relate the spin-resolved polarization densities $\{P_\sigma\}$ of an insulating state to its bosonized description, in which the fields $\phi_\sigma$ are pinned to certain values. Heuristically, a relation between $P_\sigma$ and $\langle \phi_\sigma \rangle$ can be derived by noting that, in the bosonized description, the charge densities are given by Eq.~\eqref{eq:rhos}. 
Therefore, the total charge of spin $\sigma$ at the boundary between two insulating phases where $\phi_\sigma$ is pinned to $\phi_{1,\sigma}$ and $\phi_{2,\sigma}$ is given by $\langle\int dx  \rho_\sigma(x)\rangle = \frac{1}{\pi}\Avg{\phi_{2,\sigma} - \phi_{1,\sigma}}$, where the integral is performed over a region that includes the boundary. 
On the other hand, since by definition $\partial_x P_\sigma = \langle \rho_\sigma \rangle$, we infer that $P_\sigma = \frac{1}{\pi}\langle \phi_\sigma \rangle + P_0$, where $P_0$ is a constant. 
Notice that, since $\langle \phi_\sigma \rangle$ in an insulator is defined modulo $\pi$, $P_\sigma$ is defined up to an integer~\cite{King-SmithVanderbilt1993}. 
Within our conventions, under mirror reflection $M_x$, $\phi_\sigma \rightarrow -\phi_\sigma$. 
It is therefore natural to choose $P_0=0$, such that under $M_x$, 
$P_\sigma \rightarrow -P_\sigma$.
With this choice, we express $P_\sigma$ in terms of the charge and spin fields as: 
\begin{align}
P_\uparrow &= \frac{1}{\pi\sqrt{2}}(\phi_c + \phi_s),
\nonumber\\
P_\downarrow &= \frac{1}{\pi\sqrt{2}}(\phi_c - \phi_s).
\label{eq:Psigma}
\end{align}
There are two distinct values for $P_\sigma$ that are invariant under $M_x$: $P_\sigma = 0$ and $1/2$ (mod $1$). As noted in Ref.~\cite{Efroni2017}, this implies that there are four different band insulating phases that respect the mirror symmetry. 
These phases can be accessed by tuning the axial magnetic field and (in principle) the spin-orbit coupling. 
In the presence of interactions, phases that break the mirror symmetry spontaneously are also possible. 
Below, we study the phase diagram of the system with interactions, using the polarization densities $\{P_\sigma\}$ to label the distinct phases.  
 
\section{Phase diagram}
 \begin{figure*}[ht]
 	\centering\includegraphics[width=2\columnwidth]{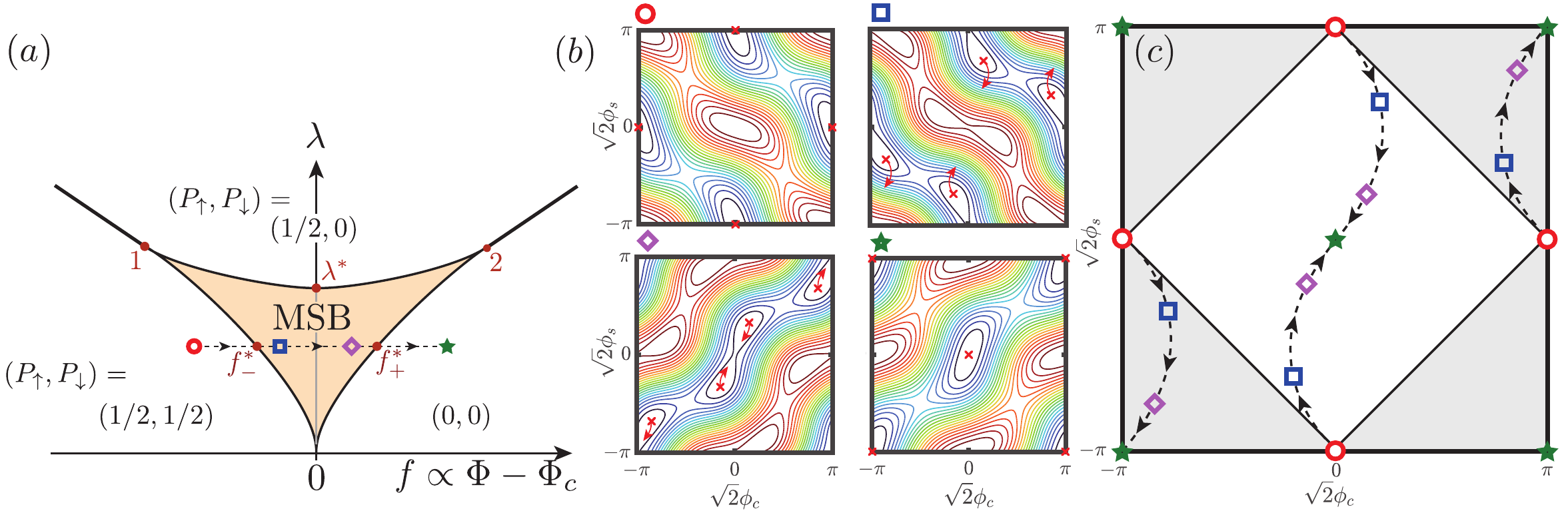}
	\caption{{Correlated insulating phases of a zigzag carbon nanotube, described the Hamiltonian in Eqs.~(\ref{eq:Hbosonized}) to (\ref{eq:BS}).
	(a) Phase diagram as a function of orbital magnetic flux $f\propto \Phi-\Phi_c$ and spin-orbit coupling, $\lambda$, at zero Zeeman field and chemical potential, $h = \mu = 0$.
	Gapped phases that respect the mirror symmetry of the CNT are labeled by corresponding values of the spin-resolved polarization densities, $P_\uparrow = \frac{1}{\pi\sqrt{2}}(\phi_c + \phi_s)$ and $P_\downarrow = \frac{1}{\pi\sqrt{2}}(\phi_c - \phi_s)$.
	Due to electron-electron interactions, a mirror symmetry breaking (MSB) phase appears over a finite range of flux and spin-orbit coupling.
	The polarization densities take nonuniversal values within the MSB phase.
	Adiabatically changing the flux along the dashed line within this phase drives spin and charge currents through the CNT, realizing a magnetoelectric effect.
	(b) Four representative snapshots of the potential landscape for the spin and charge fields, $\phi_s$ and $\phi_c$, resulting from spin-orbit coupling and the orbital magnetic flux, $H_{\lambda,f}$ in Eq.~(\ref{eq:lambdaf}), and backscattering interactions, $H_{\rm BS}$ in Eq.~(\ref{eq:BS}). Blue (red) colors represented lower (higher) potential. The parameters used in the plots are: $g_c=g_s=\lambda/3$ and $f=-2\lambda$, $-1.17\lambda$, $1.17\lambda$, $2\lambda$ for the points denoted by the red circle, blue square, purple diamond, and green star, respectively. 
	The corresponding points in the phase diagram in panel (a) are marked by the colored symbols.
	Locations of the potential minima are marked by red crosses; points separated by $(\pi, \pi)$ represent equivalent physical states.
	Red arrows indicate the direction of motion of the potential minima as the flux $f$ is scanned across the dashed line through the MSB phase, revealing how charge and spin are driven through the CNT as $f$ is varied adiabatically. (c) Motion of the minima along the dashed line in (a). Points in the white region correspond to physically distinct states. Notice that the minimum corresponding to the red circles remains at $(0,\pi)$ throughout the phase characterized by $(P_\uparrow,P_\downarrow)=(1/2,1/2)$, and similarly for the minimum corresponding to the green star.}
	}          
	\label{fig:phase_diagram}                     
\end{figure*}

As a first step, we focus on the phase diagram as a function of the two mass terms, $f$ and $\lambda$, at charge neutrality, $\mu=0$, and in the absence of a Zeeman field, $h = 0$ (the effect of the Zeeman coupling $h$ will be considered below). 
At the point $\lambda=f=0$, the coupling between the spin and charge fields in Eq.~\eqref{eq:lambdaf} vanishes, and the spin and charge sectors decouple from each other. 
In the charge sector, the Hamiltonian is of the standard sine-Gordon form. 
The term proportional to $g_c$ in Eq.~\eqref{eq:BS} is relevant for repulsive interactions ($K_c<1$). 
This term pins $2\sqrt{2}\phi_c$ to zero (mod $2\pi$) and opens a charge gap, $\Delta_c \sim \frac{v}{d}g_c^{\frac{1}{2-2K_c}}$. 
We consider the case where $\phi_c$ is pinned to zero here; it is straightforward to check that other choices (e.g., $2\sqrt{2}\phi_c=2\pi$) lead to identical physical predictions, and in fact correspond to the {\it same} ground state.
However, with $\lambda = 0$ the system is $\rm{SU}(2)$ symmetric (recall that for now, we are considering $h=0$); in this case $g_s$ is marginally irrelevant~\cite{Giamarchi2003} and the fixed point is given by $g_s=0$, $K_s=1$. 
Therefore, the spin sector is gapless at the point $\mu = h = \lambda = f = 0$.

We now consider the effect of the mass terms $\lambda$ and $f$, keeping for simplicity $h=0$. 
Turning on $f\ne 0$ opens a gap in the spectrum, whose scaling with $f$ depends on the interaction strength~\cite{LevitovTsvelik}. We may use the fact that~$\phi_c$ is pinned to zero to replace $\cos(\sqrt{2}\phi_c)$ in Eq.~\eqref{eq:lambdaf} by its expectation value (reduced from 1 due to the fluctuations of $\phi_c$). 
The remaining $\cos(\sqrt{2}\phi_s)$ is relevant, and gives rise to a spin gap $\Delta_f \sim \frac{v}{d} |f|^{2/3}$~\footnote{Unlike Ref.~\cite{LevitovTsvelik}, here we are assuming that the charge mode is gapped due to the $g_c$ (backscattering) term, resulting in a different scaling of the gap with $f$.}, up to logarithmic corrections (due to the presence of the marginally irrelevant $g_s$ term). 
The value obtained by $\phi_s$ depends on the sign of $f$: to minimize the energy, $\sqrt{2}\phi_s = 0$ (mod $2\pi$) for $f > 0$ and $\sqrt{2}\phi_s = \pi$ (mod $2\pi$) for $f < 0$.

Using the relations between $\phi_c$, $\phi_s$, and $P_\sigma$ in Eq.~\eqref{eq:Psigma}, we find that the phase obtained for $\lambda = 0$, $f \neq 0$ is described by $P_\uparrow = P_\downarrow = 0$ for $f > 0$, or $P_\uparrow = P_\downarrow = 1/2$ (mod 1) for $f < 0$.
This behavior is as expected based on the analogy to the SSH model described in Sec.~\ref{se:physical-picture} and in Ref.~\cite{Efroni2017}, in which the Wannier centers of spin-up and spin-down electrons would both be located either on the horizontal or the diagonal bonds of the zigzag CNT (see Fig.~\ref{fig:setup}a).

We now investigate the effect of spin-orbit coupling, focusing on the axis $f=0$, $\lambda\ne 0$. 
First consider the limit of very strong spin-orbit coupling, $|\lambda| \gg g_{c,s}$ (i.e., where the backscattering interactions $H_{\rm BS}$ 
can be neglected).
For $\lambda > 0$, $\phi_c$ and $\phi_s$ become pinned to values such that $\sin\left(\sqrt{2} \phi_c\right) = \sin\lp \sqrt{2}\phi_s\rp = \pm 1$, i.e., $\sqrt{2}\phi_c = \sqrt{2}\phi_s = \pi/2$ (mod $2\pi$) or $\sqrt{2}\phi_c = \sqrt{2}\phi_s = 3\pi/2$ (mod $2\pi$).
These values correspond to spin-resolved polarization densities $P_\uparrow = 1/2$, $P_\downarrow = 0$ (mod 1), as shown in Fig.~\ref{fig:setup}d; in the SSH analogy, spin-orbit coupling produces opposite dimerization patterns for spin-up and spin-down electrons~\cite{Efroni2017}.
For $\lambda < 0$, $P_\uparrow$ and $P_\downarrow$ are reversed.

Now consider the opposite limit, where a very weak spin-orbit coupling is introduced on top of the gapless state at $\mu = h = \lambda = f = 0$.
Since $\phi_c$ is pinned to zero by $H_{\rm BS}$ as described above, the $\lambda\sin \lp \sqrt{2}\phi_s\rp \sin \lp \sqrt{2} \phi_c \rp$ term in $H_{\lambda,f}$ may appear to be unimportant. 
However, to second order in $\lambda$, after integrating out the fluctuations of the massive $\phi_c$ field, we obtain a correction to $g_s$ of the form $\delta g_s \sim |A| \lambda^2$ where $A\sim K_c \log\left(g_c K_c\right)$. 
This correction breaks the SU(2) symmetry in the spin sector, and makes $g_s$ marginally relevant~\footnote{In addition to the renormalization of $g_s$, integrating out the fluctuations of $\phi_c$ also renormalizes $K_s$ upward. This increase of the spin stiffness together with the increase of $g_s$ helps pin the value of $\phi_s$.}. 
As a result, a gap $\Delta_\lambda \sim \frac{v}{d} \exp(-1/\sqrt{2g_s \delta g_s})$ opens in the spin sector, and $2\sqrt{2}\phi_s$ is pinned to a value $(2 m+ 1) \pi$, 
where $m$ is an integer. 

In order to interpret the gapped phase that arises for $f = 0$, $0<|\lambda| \ll g_{c,s}$, we examine the sublattice magnetization operator: $\mathcal{O}_s =\sum_{\sigma}\sigma (\psi_{A\sigma}^{\dagger}\psi_{A\sigma}-\psi_{B\sigma}^{\dagger}\psi_{B\sigma})\sim -\frac{4}{d}\sin\left(\sqrt{2}\phi_{s}\right)\cos\left(\sqrt{2}\phi_{c}\right)$. 
This operator is odd under mirror symmetry. In the $f=0$, $0<|\lambda|\ll 1$ phase, $\langle \mathcal{O}_s \rangle \ne 0$, and mirror symmetry is spontaneously broken. 
Notice that depending on whether $m$ is an even or an odd integer, the sign of $\Avg{\mathcal{O}_s}$ is either negative or positive, respectively.
These two possibilities reflect the two expected degenerate ground states that arise from spontaneously breaking the mirror symmetry.

To analyze how the large and small $|\lambda|$ limits are connected, notice that both the backscattering interaction term 
and the spin-orbit term 
are minimized for $\sqrt{2}\phi_s = \pi/2$ (mod $\pi$), just as in the strong spin-orbit phase described above.
Thus $\phi_s$ may remain constant throughout the parameter regime between the small and large $|\lambda|$ limits.
However, when the spin-orbit interaction is the dominant energy scale, $\sqrt{2}\phi_c$ becomes pinned to $\pm\pi/2$ (depending on whether the integer $m$ defined above is even or odd), rather than to 0 as in the small spin-orbit limit.
Therefore, due to the competition between the $g_c$ term in $H_{\rm BS}$ and the $\lambda$ term in $H_{\lambda,f}$, we expect $\sqrt{2}\phi_c$ to continuously change from 0 to $\pm \pi/2$ as $|\lambda|$ is increased from 0. 
Correspondingly, the polarization densities evolve continuously from $P_\uparrow = \pm 1/4, P_\downarrow = \mp 1/4$ to $P_\uparrow = 1/2, P_\downarrow = 0$ as we move up along the $\lambda$ axis.
In accordance with the comments about sublattice magnetization above, the polarization density values $P_\uparrow = \pm 1/4, P_\downarrow = \mp 1/4$ are indicative of a state where spin-up and spin-down electrons each exhibit unequal populations on the atomic $A$ and $B$ sublattices of the CNT. We denote the critical value of $\lambda$ at which $\sqrt{2}\phi_c$ reaches zero as $\lambda^*$ [see Fig.~\ref{fig:phase_diagram}(a)]; for $\lambda > \lambda^*$, the mirror symmetry is restored.

Next, we discuss the phase diagram with non-zero $f$ and $\lambda$, keeping $h=\mu=0$. 
As discussed above, along the axis $f=0$, $\lambda>0$ there is a gapped phase that spontaneously breaks the mirror symmetry of the system. 
We consider a cut through the phase diagram, varying $f$ at a fixed non-zero value of $\lambda<\lambda^*$ [dashed line in Fig.~\ref{fig:phase_diagram}(a)]. 
In Fig.~\ref{fig:phase_diagram}(b) we show the potential given by $H_{\lambda,f} + H_{\rm{BS}}$ [Eq.~(\ref{eq:lambdaf}) and Eq.~(\ref{eq:BS})] as a function of $\sqrt{2}\phi_c$ and $\sqrt{2}\phi_s$ at four representative points along the dashed line. 
The red crosses in the figure indicate the minima of the potential. 
Notice that points separated by $(\pm\pi,\pm\pi)$ have the same values of $P_\uparrow$ and $P_{\downarrow}$ (mod integer), and hence describe the same physical state. 

At the point in the phase diagram [Fig.~\ref{fig:phase_diagram}(a)] marked by a red circle, the corresponding potential in Fig.~\ref{fig:phase_diagram}(b) has a minimum at $(\pi,0)$, corresponding to a polarization $(P_\uparrow,P_\downarrow) = (1/2,1/2)$. 
Beyond $f=f^*_-$, this minimum splits into two minima, as shown for the point indicated by the blue square in Fig.~\ref{fig:phase_diagram}(a). 
These minima move away from each other and towards $(\phi_c,\phi_s) = (0,0)$ as $f$ increases [see arrows indicating the direction of motion for the potential minima, and the potential corresponding to the point marked by purple diamond in Fig.~\ref{fig:phase_diagram}(b)], until they merge at $(\phi_c,\phi_s) = (0,0)$ when $f = f_+^*$. 
Beyond this point, the minimum is at the origin, corresponding to a polarization $(P_\uparrow,P_\downarrow) = (0,0)$ (see potential at the point marked by the green star). Fig.~\ref{fig:phase_diagram}(c) shows the motion of the potential minima along the dashed line in Fig.~\ref{fig:phase_diagram}(a). 

Within the mirror symmetry breaking (MSB) phase [filled region in Fig.~\ref{fig:phase_diagram}(a)], both $\phi_c$ and $\phi_s$ change continuously as $f$ is varied, corresponding to changes in the polarization densities $P_{\uparrow,\downarrow}$. 
Therefore, an adiabatic change in the axial field within this phase causes spin and charge currents to flow along the nanotube. 
As $f$ changes from $f^*_-$ to $f^*_+$, $\sqrt{2}\phi_s$ changes by $\pm\pi$. 
This change of $\phi_s$ corresponds to one net electron spin being pumped across the system. 
The direction of the pumping depends on which of the two degenerate grounds states the system is in.
Additionally, $\phi_c$ also changes as $f$ varies from $f^*_-$ to $f^*_+$ [see potentials corresponding to the blue square and purple diamond, displayed in Fig.~\ref{fig:phase_diagram}(b) and Fig.~\ref{fig:phase_diagram}(c)]. 
These changes of $\phi_c$ imply that the charge polarization density changes as the magnetic field is varied, realizing a magneto-electric effect. 
Unlike $\phi_s$, the change of $\phi_c$ is non-monotonic, with no net charge being pumped as $f$ changes from $f_-^*$ to $f_+^*$. 
The maximum change in the charge polarization is obtained at an intermediate value of the flux within the MSB phase. 
Its magnitude is non-universal, depending on microscopic parameters such as $\lambda$, $g_c$, and the Luttinger parameter $K_c$.

For a sufficiently large $\lambda$, the MSB phase disappears for all values of $f$. Instead, there is a direct transition from the $(P_\uparrow,P_\downarrow) = (1/2,1/2)$ phase to the $(1/2,0)$ phase, and another transition from the  $(1/2,0)$ phase to the $(0,0)$ phase, as in the non-interacting case~\cite{Efroni2017}. Across these transitions, the charge polarization density changes discontinuously by 1/2. 
Along the transition lines [the thick black lines emanating from the points 1 and 2 in Fig.~\ref{fig:phase_diagram}(a)] the interaction terms in Eq.~\eqref{eq:BS} are irrelevant. 
At points 1 and 2, the interaction terms are marginal. 
Between these points, each transition line splits in two, and the MSB phase is formed. 
The boundaries of the MSB phase (except the points 1, 2, and the origin) are likely to be of the 1+1 dimensional Ising universality class, since along these lines the $\mathbb{Z}_2$ mirror symmetry is broken. 

To examine the effect of the Zeeman field $h$, we start from the point $\lambda=f=\mu=0$. 
At this point, the Zeeman field leaves the spin sector gapless, and induces a finite spin magnetization. 
To see this, we write $\phi_s(x) = \tilde{\phi}_s(x) + \frac{\sqrt{2}h K_s}{v_s} x$. 
Then, in terms of $\tilde{\phi}_s$, the $h$ term in Eq.~\eqref{eq:Hz} disappears, and the cosine in the $g_s$ term [Eq.~(\ref{eq:BS})] becomes spatially modulated; the effect of the cosine term is thus suppressed by the Zeeman field. Additionally, there is a constant term in the energy density,~$\frac{K_s}{\pi v_s} h^2$. 
Turning on $f$ and $\lambda$, a gapless phase with a finite spin magnetization density appears in a region around the point $f=\lambda=0$, whose extent depends on $h$.
The critical lines at large $\lambda$ beyond the points 1 and 2 [thick black lines in Fig. \ref{fig:phase_diagram}(a)] similarly expand into gapless regions upon increasing $h$. 
Away from points 1, 2 and the origin $\lambda=f=0$, the phase diagram with a small non-zero $h$ is similar to that shown in Fig.~\ref{fig:phase_diagram}(a). (For brevity, we do not plot the $h\ne 0$ phase diagram here.)
In particular, the existence of a MSB phase is stable for sufficiently small $h$. 
Similar considerations can be applied for the phase diagram at a non-zero chemical potential,~$\mu$.


 \section{Discussion}
 In this work we described how electron-electron interactions in zigzag carbon nanotubes lead to the formation of correlated insulating phases. The interactions prevent the magnetic field driven gap closing expected based on the single particle band structure.
 In particular, over a range of values of spin-orbit coupling and magnetic flux, we predict that the ground state of the carbon nanotube spontaneously breaks the mirror symmetry of the system.
 Within this MSB phase, adiabatic changes of the flux drive charge and spin currents along the nanotube.
 
 In experiments, the spin-orbit coupling is fixed for a given nanotube.
 Therefore, the applied flux (along with the Zeeman field and chemical potential) is the primary tool for exploring the phase diagram.
 To assess whether or not the MSB phase can be realized, we must compare the value of the spin-orbit coupling to the critical value~$\lambda^*$ above which the symmetry breaking phase disappears.
 The value of $\lambda^*$ is determined by the competition between the spin-orbit coupling term $\lambda \sin \left( \sqrt{2} \phi_s \right) \sin \left( \sqrt{2} \phi_c \right)$ in $H_{\lambda, f}$, Eq.~(\ref{eq:lambdaf}), and the $g_c \cos \left( 2\sqrt{2}\phi_c\right)$ term in $H_{\rm BS}$, Eq.~(\ref{eq:BS}). 
 Experimentally, spin-orbit coupling values of up to a few meV have been reported in CNT quantum dots~\cite{IlaniKuemmeth2008, Steele2013, LairdRMP}.
 In contrast, we estimate the bare value of $\frac{v g_c}{d} = \frac{U_-}{2\pi d} \approx \frac{e^2}{d} $ [see text below Eq.~(\ref{eq:BS})] to be of the order of a few hundred meV. Therefore, it is likely that in experiments, $\lambda<\lambda^*$. However, whether the MSB phase is realized also depends on the strength of the Zeeman term $h$, which is not independent of the orbital magnetic flux $\Phi$ (notice that $h\ne 0$ when $\Phi = \Phi_c$). 
Therefore, as $\Phi$ is swept across $\Phi_c$, the system either enters the MSB phase, or a phase with a non-zero spin magnetization and gapless spin excitations.  
 In either case, the charge gap remains open, in agreement with the experimental observation \cite{Bockrath2009}. 
 In view of future experiments, closing of the spin gap near $\Phi = \Phi_c$ could be detected via the associated enhancement of magnetic field fluctuations.

\section{Acknowledgements}

This work was supported by the DFG (CRC / Transregio 183, EI 519/7-1). 
The research at WIS was supported by the European Union’s Horizon 2020 research and innovation programme
[grant agreements LEGOTOP, No. 788715 (YO) and HQMAT, No. 817799 (EB)], the BSF and NSF (2018643), and the ISF
Quantum Science and Technology (2074/19).
MR is grateful to the Villum Foundation, the University of Washington College of Arts and Sciences and the Kenneth K. Young Memorial Professorship for support.

 

\bibliography{ref}
\end{document}